\documentclass[amssymb,amsfonts,aps,prl,twocolumn,showpacs]{revtex4}

\usepackage{graphicx}
\usepackage{dcolumn}
\usepackage{bm}

\begin{document}

\title{In-gap impurity states as the hallmark of the Quantum Spin Hall phase}

\author{ J. W. Gonz\'alez$^1$, J. Fern\'andez-Rossier$^{1,\,2}$.}
\affiliation{
(1)  International Iberian Nanotechnology Laboratory - INL,
Av. Mestre Jos\'e Veiga, 4715-330 Braga, Portugal
\\
(2) Departamento de F{\'i}sica Aplicada, Universidad de Alicante, San
Vicente del Raspeig, Spain}

\date{\today} 

\begin{abstract} 
We study the  different response  to  an impurity  of the two 
topologically different phases shown by a two dimensional insulator with time reversal 
symmetry, namely, the Quantum Spin Hall and the normal phase. We consider the case  of graphene as a toy model that features the   two phases driven,  respectively, by
intrinsic spin-orbit coupling and  inversion symmetry breaking.  We find that  strictly normalizable in-gap impurity states only 
occur in the Quantum Spin Hall phase  and  
 carry dissipationless current  whose quirality is determined by the spin and 
pseudospin of the residing electron.   Our results imply that topological order
can be unveiled by  local probes of defect states.
\end{abstract}

\maketitle

The intrinsic properties of an electron gas are revealed in the way it reacts to the 
presence of a localized perturbation. 
In metals,   the period of Friedel oscillations   provides information about the Fermi surface \cite{Friedel}.   In semiconductors, 
the binding energy of shallow acceptors and donor  states  depends 
on the effective mass and dielectric constant of the 
host material \cite{Hydrogenic}. In superconductors, both the 
impurity induced modulations of the density of states \cite{Byers93}
and the presence of zero energy or mid-gap states \cite{Hu_PRL1994},   reveal the symmetry of their order parameter.  Here we 
address the fundamental question of whether two dimensional topological insulators  
react to a localized perturbation in a way different  from conventional insulators.  
 
Topological insulators \cite{Hasan-Kane-2010} have a bulk bandgap like a normal insulator 
but have protected conducting states on their edges and  surfaces. This bulk-boundary 
correspondence has been  rigorously established \cite{Halperin82,Zhang06} in  boundaries that preserve the translational invariance in at least  one 
dimension. 
In two dimensional topological insulators,  edge state are expected to present quantized conductance \cite{Kane-Mele1,Kane-Mele2,Murakami2011} which can be used to unveil 
the existence of bulk topological order  \cite{Konig2007}.
More recently, the bulk-boundary correspondence has been extended to the 
case of topological defects that lead to Hamiltonians ${\cal H}(\mathbf{k},\mathbf{r})$ 
that vary slowly with adiabatic parameters $\mathbf{r}$ surrounding the defect 
\cite{Teo-Kane}.

Here we explore the electronic structure of a   two dimensional insulator, that can be either in the 
Quantum Spin Hall  (QSH) or in the normal phase,  in the neighborhood of
an isoelectronic impurity that creates a repulsive  short range potential. 
We find  that only in the QSH phase in-gap states appear and have exotic electronic properties:
they carry  non-dissipative spin current. 
Our finding provides an alternative way to detect topological order, using local probes sensitive to  
the density of states in the neighborhood of the  impurities.

We use  gapped graphene as a toy model for two dimensional topological 
insulators \cite{Kane-Mele1,Kane-Mele2}.  
The electronic properties of graphene are intimately related to the structure of the 
honeycomb lattice,   formed by two interpenetrating triangular 
sublattices,  $A$ and $B$, related by inversion symmetry, which define a pseudospin degree 
of freedom that we denote with the operator $\tau_z$. 
There are two ways to open a gap  in graphene preserving the size of its two atom minimal 
unit cell. Both  lead to interesting electronic phases.  
A conventional gap opens in graphene when a sublattice symmetry breaking potential, 
$\frac{\Delta_0}{2}\tau_z$, is included in the Hamiltonian.  
This gap entails peculiar electronic properties:  the two valleys carry orbital 
currents of opposite sign  \cite{Xiao_PRL2007}.       
When the gap is opened by intrinsic spin orbit coupling, as described with the 
second-neighbour spin dependent hopping proposed by Kane-Mele,  graphene is in the QSH phase.

The Kane-Mele \cite{Kane-Mele1,Kane-Mele2} tight-binding Hamiltonian  ${\cal H}_0$ 
describes electrons in a hexagonal lattice, with first neighbor hopping $t$, 
spin-dependent second neighbor hopping $t_{so}$ and  the $\frac{\Delta_0}{2}\tau_z$ term.  This model commutes with $S_z$, the spin projection perpendicular to the 
graphene plane, and each spin sector is identical to the Haldane model \cite{Haldane88} 
for spinless fermions. 
We consider the effect of a substitutional isoelectronic impurity, described 
with a  single-site potential,
  in  an otherwise boundless and 
perfect  gapped   system:  
\begin{equation}
{\cal H}={\cal H}_0+ V_0 \sum_s |0s\rangle\langle 0 s\vert , 
\label{HV} 
\end{equation} 
The strength of the spin-independent  impurity potential is $V_0$, and acts 
only on the atom $0$ of the $A$ sublattice.

For each spin channel $s$,  the crystal Hamiltonian ${\cal H}_0$ can be written in the form of a 
2x2  matrix in the sublattice basis: 
\begin{equation} \label{H} 
\mathcal{H}_{\mathbf{k}}^s = 
t\left[ f(\mathbf{k})\tau^+  + f^*(\mathbf{k})\tau ^ -\right] 
+ \frac{\Delta_0 }{2} \tau_z  +
s t_{so}\, g (\mathbf{k})\tau_z  ,
\end{equation}
where $\vec{\tau}=(\tau_x,\tau_y,\tau_z)$ are the Pauli matrices in the sublattice space, 
$\tau^{\pm}=\tau_x\pm i\tau_y$, $s=\pm1$ are the eigenvalues of  the spin operator $S_z$. 
$f(\mathbf{k})$ and $g(\mathbf{k})$ are the usual functions that sum the Bloch 
phase over the nearest and path-dependent next-nearest neighbors \cite{Kane-Mele2}.

The Bloch Hamiltonian is identical to that of a (pseudo)spin  $\vec{\tau}$ 
in an effective field: $\mathcal{H}_{\mathbf{k}}^s= \vec{h}^s_\mathbf{k} \cdot 
\vec{\tau}$, where 
$$\vec{h}^s_\mathbf{k}= 
\left[t f(\mathbf{k}) + t f^*(\mathbf{k}),\, t f(\mathbf{k}) - t f^*(\mathbf{k}), 
\frac{\Delta_0 + t_{so}\, g (\mathbf{k})}{2}
\right]. $$

\begin{figure}[ht!]
\includegraphics[clip,width=0.5\textwidth,angle=0,clip]{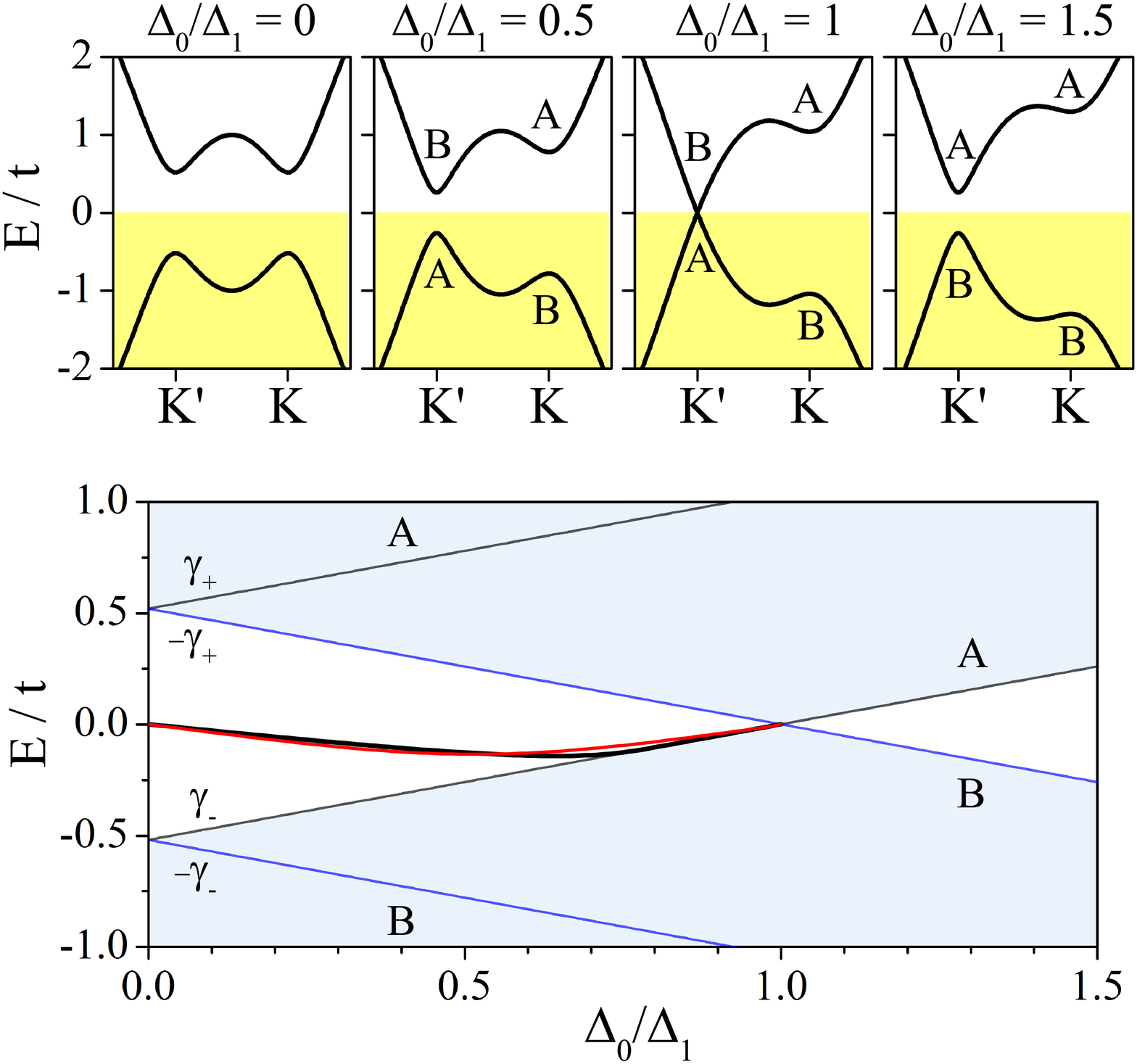} 
\caption{(Color online) Top panel: energy bands, for  the $s=+$ spin channel, for 
$t_{so}=0.1t$, and 4 values of $\Delta_0$. 
Bottom panel: evolution of bands at $K$ and $K'$ and in-gap impurity energy $E_b(\Delta_0)$
  calculated both  with the lattice model (black) and the 
continuum model (red).}
\label{Fig1:band} \end{figure}

The energy bands for the model are given by
$\epsilon_{\nu,s}(\mathbf{k})= \pm \epsilon^s_{\mathbf{k}}$,
where $\nu=\pm$ labels the two bands (per spin channel) and
$\epsilon^s_{\mathbf{k}}\equiv |\vec{h}^s_\mathbf{k}|$.
In the top panel of Fig. \ref{Fig1:band} we plot them for a given value of $s$, 
along the line that joins the Dirac points  $K$ and $K'$,
all of them with the same values of $t$ and  $t_{so}=0.1 t$ and different 
values of $\Delta_0$.  
Time reversal invariance ensures that, for the opposite spin orientations, we have 
$\epsilon_{\nu,s}(\mathbf{k})=\epsilon_{\nu,\overline{s}}(-\mathbf{k})$.
The $K$ and $K'$  points  define the so called valley  index $\sigma=\pm 1$. 
These points are special because the in-plane components of the 
effective field $\vec{h}^s_\mathbf{k}$ vanish, making the energy splitting between the 
bands minimal.  
At $K$ and $K'$ we have $\vec{h}^s_\mathbf{k}= (0,0,\gamma_{s\sigma})$ with: 
\begin{equation}
  \gamma_{s\sigma}=\frac{1}{2}\left( \Delta_0 +   s \sigma \Delta_1 \right),
  \end{equation} 
where $s\sigma$ can take only two values $s\sigma=\pm 1$ and 
$\Delta_1= 6\sqrt{3}t_{so}$.  Thus, at the Dirac point the wave functions 
have a well defined sublattice ($\tau_z$) character. The  effective field at the Dirac 
points, $\pm \gamma_{s\sigma}$, defines the top of the valence band and bottom of the 
conduction  band at each valley. We plot them in the low panel of Fig. \ref{Fig1:band}.  
  
In the transition from the QSH phase,  with  $\gamma_-<0$ to the normal phase, 
with   $\gamma_->0$,   the system closes the gap ($\gamma_-=0$),  reflecting the impossibility of deforming adiabatically one phase into the other \cite{Kane-Mele1,Kane-Mele2}. 
In the QSH phase,  the energy splitting is finite at both valleys,  
the orientation of the  effective field $\vec{h}^s_\mathbf{k}$ is opposite at $K$ 
and $K'$, for a given $s$. In this phase,  the model presents 
topologically protected edge states inside the gap, making the edge metallic. 
At the normal phase, with   $\gamma_->0$  the orientation of the effective  field  $\vec{h}^s_\mathbf{k}$ is now the same at both valleys and a gap opens in the edge states.

We now explore the possible appearance  of in-gap states in the two insulating phases of 
the model in the presence of the impurity potential, as described by Eq. (\ref{HV}). This is different from previous works that have studied the influence of impurities on the conducting surface states of 3D topological insulators \cite{Biswas2010,Balatsky2012}.
We address our problem using the T-matrix formalism \cite{Economou}.  
We define the Green function operators, $\mathcal{G}(z)\equiv 
\left(z\mathbb{I}-\mathcal{H}\right)^{-1}$, and 
$\mathcal{G}_0(z)\equiv \left(z\mathbb{I}-\mathcal{H}_0\right)^{-1}$ where $z$ is a 
complex number and $\mathbb{I}$ is the unit matrix in the Hilbert space of the infinite 
lattice. 
A closed expression for the  Green function   $\mathcal{G}_0(z)$ can be readily 
obtained in terms of the eigenstates of $\mathcal{H}_{\mathbf{k}}^s$. 
For the on-site spin-independent potential,
the complete Green function is related to $\mathcal{G}_0(z)$  by:
\begin{equation}\label{Te}
\mathcal{G}(z) = \mathcal{G}^0(z) 
+ \mathcal{G}^0(z) \vert 0 \rangle \frac{ V_0 }{1-V_0 \mathcal{G}^0_{0,0}(z)}  \langle 0 
\vert \mathcal{G}^0(z),
\end{equation}
where $ \mathcal{G}^0_{0,0}(z)$ is the unperturbed Green function diagonal matrix element 
in the atomic representation and $0$ is the atom at which the perturbation is located. 

The  appearance of bound states with energy $E_b$ would be given by the 
existence   of  poles of $\mathcal{G} $ in the band-gap defined by 
$\mathcal{H}_{\mathbf{k}}^s$.  Thus, we have to solve the equation
\begin{equation}\label{Ep_T}
\mathcal{G}^0_{0,0} (E_b) = \frac{1}{V_0}.
\end{equation}  
In order to have a closed expression for $\mathcal{G}^0_{0,0} (E_b) $ we use the Lehman 
representation and project over the site representation. 
Any given site in the lattice $i$ can be identified by its unit cell $\vec{R}_i$ and the 
sublattice  $\tau_i=A,B$.  
The unperturbed Green function is written in terms of the eigenstates of the crystal 
Hamiltonian
 $$|\nu,\mathbf{k},s\rangle=\frac{1}{\sqrt N} \sum_{\vec{R},\tau} 
 e^{i\mathbf{k}\cdot\mathbf{R}} 
 {\cal U}^s_{\nu,\mathbf{k},\tau} |\vec{R}\tau,s\rangle,$$ where $N$ is the number of unit 
cells of the crystal,  and ${\cal U}^s_{\nu,\mathbf{k},\tau}$ are the components of the 
eigenstates of the  Bloch Hamiltonian in Eq. (\ref{H}).  If we 
express the effective field in spherical coordinates as,
 $\vec{h}^s_\mathbf{k} =  \epsilon^s_\mathbf{k}  \left( \sin \theta^s_\mathbf{k}  \cos 
\varphi^s_\mathbf{k},\,\sin \theta^s_\mathbf{k}  \sin 
\varphi^s_\mathbf{k},\,\cos \theta^s_\mathbf{k}  \right) $,
the corresponding wave functions for the $\nu=\pm$ bands read:
\begin{eqnarray} \label{autov3}
 {\cal U}^s_{-,\mathbf{k},A}&=&\sin \left(  \frac{\theta}{2} \right) 
e^{-i \varphi}, \,\, \,\,
  {\cal U}^s_{-,\mathbf{k},B}= - \cos \left( \frac{\theta}{2} \right),\nonumber\\
{\cal U}^s_{+,\mathbf{k},A}&= &\cos \left( \frac{\theta}{2} \right) e^{-i \varphi},  \,\, 
\,\,{\cal U}^s_{+,\mathbf{k},B}=
\sin \left( \frac{\theta}{2} \right),
\end{eqnarray}
where   we  have omitted the subscripts from $\theta$ and $\varphi$  for the sake of 
clarity.  The unperturbed Green function matrix can be written as: 
 \begin{equation} \label{G}
 \mathcal{G}^0_{ij} (z) =\frac{1}{N} \sum_{\mathbf{k},\nu} \frac{e^{\mathbf{k}\cdot \left( 
 \mathbf{R}_i - \mathbf{R_j} \right) }}{z-\varepsilon_{\mathbf{k},\nu}}
\left({\cal U}^s_{\nu,\mathbf{k},\tau_i}\right)^*
{\cal U}^s_{\nu,\mathbf{k},\tau_j.}\end{equation}
Using this expression in combination with Eq. (\ref{autov3}), we can recast the Eq. 
(\ref{Ep_T}) as
\begin{equation} \label{GAA_V0}
\frac{1}{V_0} =\frac{1}{2N} \sum_\mathbf{k}
 \left[ \frac{ 1-\vec{n}^s_{\mathbf{k}} \cdot\hat{z}}{E_b 
+\epsilon^s_\mathbf{k}}  +
\frac{ 1+\vec{n}^s_{\mathbf{k}}\cdot\hat{z}}
{E_b-\epsilon^s_\mathbf{k}}\right] ,
\end{equation}
where $\vec{n}^s_{\mathbf{k}}=\vec{h}^s_{\mathbf{k}}/\epsilon^s_{\mathbf{k}}$.
Importantly,  the Green function $\mathcal{G}^0_{00} (z)$ is expressed as a sum over the 
whole Brillouin zone
of a function that depends on  the projection of the effective field vector and, as such, 
it contains information  of the topology of the Bloch states, which is a necessary 
condition to expect a relation between the solutions of Eq. (\ref{GAA_V0}) and the 
topological order in the system.
 
The Eq. (\ref{GAA_V0}) can be analytically solved in the strong coupling limit 
$V_0^{-1}=0$,  in two cases. For $\Delta_0=0$,  the repulsive potential yields a mid-gap 
state $E_b=0$ whose properties we discuss below.  For $\Delta_1=0$, there is a solution 
with $E_b=-\Delta_0/2$, 
in agreement with a general result \cite{Soriano-JFR2012}. 
However, this solution is degenerate with the top of the valence band and it is a 
resonance rather than  an actual in-gap state. 
The interpolation between these two limits is obtained by the numerical solution of 
Eq. (\ref{GAA_V0}) and is shown in Fig. \ref{Fig1:band} for $V_0=10^6t$. 
Bound states are only found when $V_0>t$ and,  interestingly, only 
when $\Delta_0<\Delta_1$, i.e., in the topological insulator phase.  This is the main 
result of the paper: we find that a local impurity can bind an in-gap state only in the 
QSH phase. 
 
In order to obtain some analytical insight of the one to one relation between the 
existence of in-gap states and the topological phase, we have worked out Eq. (\ref{G}) in 
the continuum limit, in which only states close to the two Dirac points are included. 
Their crystal Hamiltonian is then given by  
\begin{equation}
{\cal H}_0(\vec{k},\sigma,s) \!=\! \hbar v_F \left(  k_x \tau_x + i \sigma k_y \tau_y 
\right) + \frac{\Delta_0}{2} \tau_z + \frac{\Delta_1}{2} s\sigma \tau_z.
\end{equation}
For this model it is possible to obtain a closed analytical expression for Eq. 
(\ref{GAA_V0}): 
\begin{equation} \label{V0}
\frac{1}{V_0}
\! =\! \frac{a^2}{2\pi\!\left( \hbar v_F \right)^2} \! \sum_{s \tau=\pm}  
 \!\! \left(\gamma_{s\tau} \!- \! E_b \right)
\, \log \left[1 + \frac{ \varepsilon_c^2}{\gamma_{s\tau}^2 - E_b^2} \right] ,
\label{Epcont}
\end{equation}
where $\epsilon_c$ is the cutoff energy \cite{Inglot2011}. 
The sum over $s\tau$ reflects, for a given spin orientation,  the contributions  coming 
from both valleys.   
 In the strong coupling limit, $V_0^{-1}$ is negligible and the existence of an in-gap 
solution of Eq. (\ref{Epcont}) requires that the right hand side sum vanishes. 
Since $\gamma_{+}-E_b$ is always positive,  $E_b$ must satisfy $\gamma_-<E_b$, which, by definition of in-gap state,  is only satisfied in the QSH phase and not in the normal phase (see lower panel of Fig. \ref{Fig1:band}).  Thus, the continuum model also has the one on one relation between topological order and the emergence of in-gap impurity states. 
The numerical solution of Eq. \ref{Epcont}, for $V_0^{-1}=0$ is shown in Fig. \ref{Fig1:band}.  

\begin{figure}[t!]
\includegraphics[clip,width=0.49\textwidth,angle=0,clip]{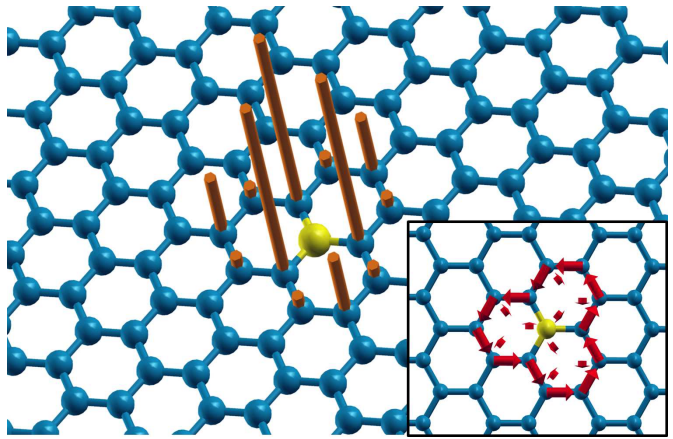} 
\caption{(Color online) Local density of states as a bar in a exponential scale.
The in-set shows the impurity-induced current between links for the state at
$E_b$. The impurity site is shown in yellow.}
\label{fig2:Curr}
\end{figure}

Now we address the electronic properties of the topological in-gap 
impurity states.  The Green's function formalism provides a closed expression   
for the in-gap wave function in terms  of their binding energy $E_b$, obtained 
from Eq. (\ref{G}), and the unperturbed Green function \cite{Economou}:
\begin{equation}\label{wf_anali}
\vert \phi_b \rangle = 
\left(-\frac{d\mathcal{G}^0_{00}(E_b)}{dE} \right)^{-\frac{1}{2}}
\sum_i 
\mathcal{G}^0_{i,0} (E_b)
\vert i \rangle.
\end{equation}
We consider the case $\Delta_0=0$ and $V_0^{-1}=10^{-6}t$. 
Expectedly, the in-gap wave function  is localized around the impurity site, 
as shown in Fig. \ref{fig2:Curr}. Interestingly, the in-gap states 
is not fully   sublattice polarized, in contrast with zero energy states in bipartite Hamiltonians \cite{Palacios2008}. Thus, we have 
$\langle \phi_b |\sigma_z|\phi_b\rangle\simeq 0.56$,
significantly below $1$.
We also find, analytically,  that   the 
wave function in the impurity sublattice   is  purely imaginary, while for the other 
sublattice is  purely real.
 In the reciprocal space it is also true that the wave function has 
unequal weight on both valleys, but is not fully valley polarized.  Both the incomplete sublattice and valley polarizations
are quantitatively different from spin-filter edge states in zigzag ribbons for the same 
model, which are fully sublattice and valley polarized.   

In contrast, the most salient 
feature of the edge states is also present for the topological in-gap impurity states: 
they carry current, which is quite unusual for a localized state.   
The current operator is defined at the bonds
 of  the tight-binding Hamiltonian 
imposing the continuity equation.  For a given pair of sites $n$ and $m$ in the lattice, 
the current operator reads  \cite{Soriano-JFR10}
$ \hat{J}_{nm}= j_{nm} |n\rangle\langle m| - j_{mn} |m\rangle\langle n|, $ where $j_{mn} 
= et_{nm}/i\hbar$ and $t_{nm}$ is the Hamiltonian matrix element $\langle n|{\cal H}_0|m\rangle$. 
Thus,  the current carried by bound states reads
\begin{equation}
I^{\rm bound}_{nm}= \frac{e}{\hbar} \mathrm{Im} \left[  t_{nm} \phi^*_b(m)
\phi_{b}(n)\right].
\end{equation}
The map of the current is shown, for a given spin,  in the inset of Fig. 
\ref{fig2:Curr}, 
for $t_{so}=0.1 t$, $\Delta_0=0$, $V_0=10^{-6}t$.  Although  the in-gap state is more localized in the $B$ sublattice,
opposite to the impurity site, the current is larger in the $AB$ bonds than in the $BB$ bonds
 because $t=10 t_{so}$.  
For the opposite spin, the current flow changes sign  so that the net current 
is zero but the spin-current is not.  We  have also verified that,  when the impurity 
site is in the other sublattice,  the
current flow is inverted. Thus, the in-gap states have dissipationless spin-currents
 whose chirality is determined by the sublattice at which the impurity resides.

\begin{figure}[hbt!]
\includegraphics[clip,width=0.49\textwidth,angle=0,clip]{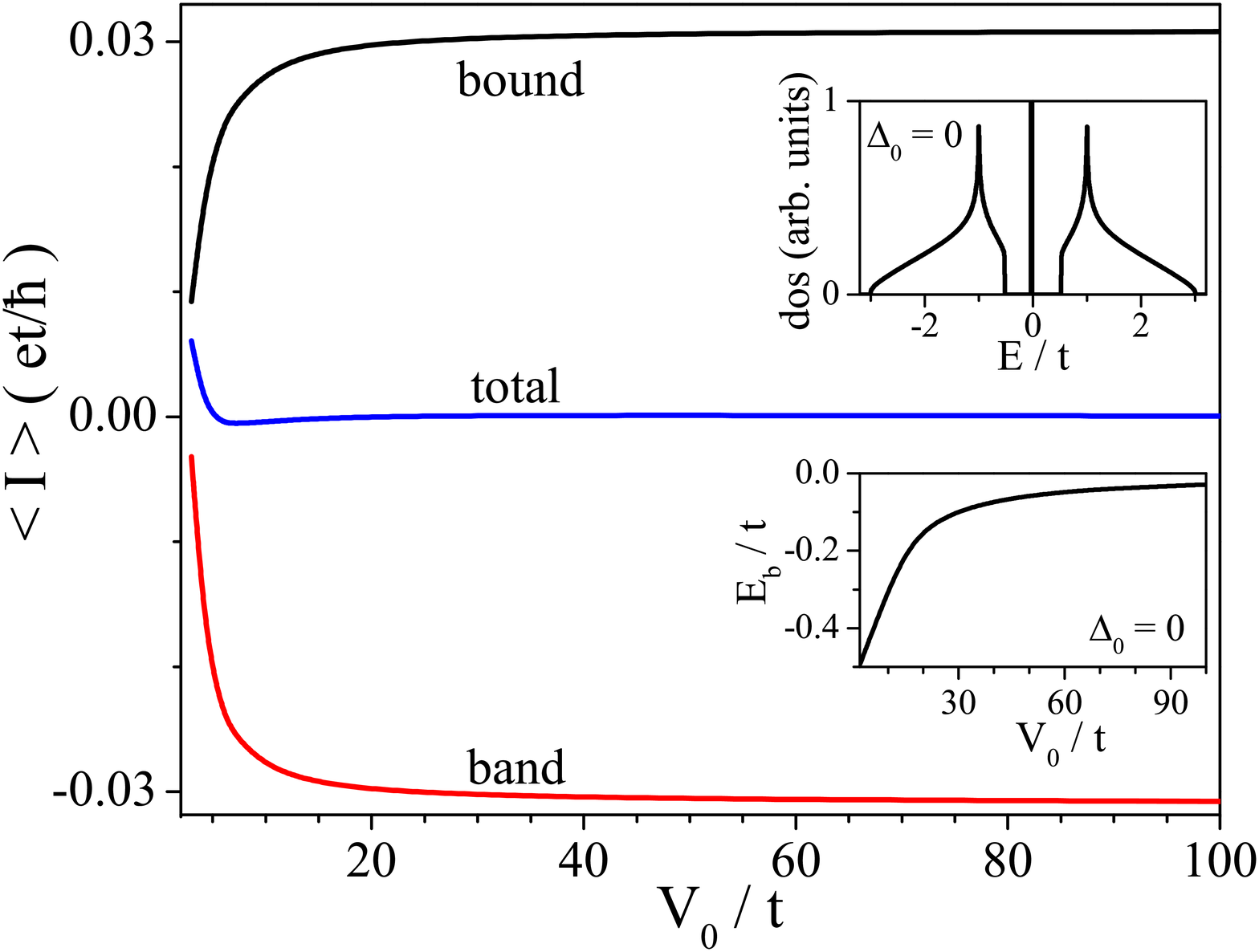}
\caption{(Color online) Bound, band and total  current in an $AB$ bond , as a function of $V_0$. Upper inset: density of states on $B$ sublattice.  Lower inset: evolution of $E_b(V_0)$ for $\Delta_0=0$.  
}
\label{Fig3:Imax}
\end{figure}

The observable current probed experimentally  would be given by the contribution of all 
occupied states, which includes both the in-gap state and the valence band, whose density of states are shown in the upper inset of Fig. \ref{Fig3:Imax}.  The contribution to the 
current from the band  states reads
\begin{equation}
I^{\rm band}_{nm} \! \equiv \! \! \frac{-1}{\pi}\!\! 
\int\limits_{-\infty}^{\gamma_-} 
\!\!  \mathrm{Im} \left[ {\cal G}_{nm}(E)j_{mn} - {\cal G}_{mn}(E)j_{nm}\right] 
\mathrm{d}E.
\end{equation}
where  ${\cal G}$ is  the full Green function whose closed form  is given in Eq. 
(\ref{Te}).
Thus, there are two contributions to the current, one given by the band states and the 
other given by the in-gap states.  We plot both them in Fig. \ref{Fig3:Imax}, for $\Delta_0=0$, 
as a function of the impurity strength, $V_0$.  They have opposite signs and, in the 
strong $V_0$ limit, cancel each other.  The evolution of $E_b(V_0)$ is shown in the lower inset.  In that limit the impurity bound state becomes a 
mid-gap state $E_b=0$ and the spectrum recovers electron-hole symmetry, for which ground 
state  currents are not possible \cite{Wu2011}.  The cancellation of the edge current due 
to the contribution of the bulk states takes also place in the case of ribbons.
However, a finite non-zero  spin-current is obtained in our case  for a wide range of 
$V_0$ close to what it is expected for absorption of atomic hydrogen in graphene.  

It must be stressed that, in contrast to in-gap states in the domain wall of 
polyacetylene,  the impurity in-gap states do not give rise to charge fractionalization.  
This can be understood as follows. Together with the in-gap solution of Eq. (\ref{Ep_T}) 
where $E_b\simeq V^{-1}_0$, there is always a second solution with energy $E_b\simeq V_0$. 
Thus, there are two states outside of the bands, and not only one as in the case of 
Su-Schrieffer-Heeger solitons \cite{Schrieffer}, so that both bands lose a complete state. 

In conclusion, we have studied the electronic properties of the reaction of a  
two dimensional insulator, as given by the  the Kane-Mele model, to a strong on-site 
impurity
potential.  We have shown that in-gap impurity states  appear in the case 
topologically non-trivial or Quantum Spin Hall phase \cite{Kane-Mele1,Kane-Mele2}.  
Vacancy induced in-gap states have been also predicted for  a two-dimensional topological insulator described with a modified Dirac equation \cite{Shen2011}. Here we have shown the one-on-one relation between the topological order and the existence of the in-gap states in the context of the Kane-Mele lattice model. 
We have found that these  topological in-gap impurity states carry a net spin-current which, 
due to the single occupancy of the state, is upgraded to a net current or orbital 
magnetization.  Therefore, we propose that the topological nature of this type of 
insulators can be stablished by local probes of the electronic properties of these 
defects,  as opposed to the highly non-local probes required to test quantized 
transport of the edge states.   
Further work should determine if our results can 
be extended to higher dimensions and to other models for topological insulators 
in two dimensions.

This work has been financially supported by MEC-Spain (Grant Nos. FIS2010-21883-C02-01  
and CONSOLIDER CSD2007-0010). 
We are indebted to  A. Balatsky  for fruitful discussions.

\end{document}